\documentclass[reprint,superscriptaddress,amssymb,amsmath,aps,showpacs,10pt,floatfix,prl,longbibliography]{revtex4-2}

\def\KVS{KV$_3$Sb$_5$}
\def\RVS{RbV$_3$Sb$_5$}
\def\RTB{RbTi$_3$Bi$_5$}
\def\ATB{$A$Ti$_3$Bi$_5$}
\def\CTB{CsTi$_3$Bi$_5$}
\def\CVS{CsV$_3$Sb$_5$}
\def\AVS{$A$V$_3$Sb$_5$}
\def\Fe3Sn2{Fe$_3$Sn$_2$}
\def\cm{cm$^{-1}$}

\usepackage{graphicx}%
\usepackage{color}
\usepackage{epstopdf}
\usepackage{placeins}
\usepackage{amssymb}
\usepackage{amsmath}
\usepackage{amsfonts}
\usepackage[note-name=, use-sort-key = false]{notes2bib}

\usepackage{xcolor}

\usepackage{color}
\usepackage[colorlinks,bookmarks=false,citecolor=darkblue,linkcolor=red,urlcolor=blue]{hyperref} 
\definecolor{darkred}{rgb}{0.7,0.0,0.0}

\definecolor{darkblue}{rgb}{0,0.02,0.45}

\definecolor{darkgreen}{rgb}{0.02,0.45,0.0}

\definecolor{violet}{rgb}{0.8,0.2,0.6}

\begin{document}
\title{Interplay of $d$- and $p$-states in \RTB\ and \CTB\ flat-band kagome metals }

\author{M. Wenzel}
\email{maxim.wenzel@pi1.physik.uni-stuttgart.de}
\affiliation{1. Physikalisches Institut, Universit{\"a}t Stuttgart, 70569 
Stuttgart, Germany}

\author{E. Uykur}
\affiliation{Helmholtz-Zentrum Dresden-Rossendorf, Institute of Ion Beam Physics and Materials Research, 01328 Dresden, Germany}

\author{A. A. Tsirlin}
\affiliation{Felix Bloch Institute for Solid-State Physics, Leipzig University, 04103 Leipzig, Germany}

\author{A. N. Capa Salinas}
\affiliation{Materials Department and California Nanosystems Institute, University of California Santa Barbara, Santa Barbara, CA, 93106, USA}

\author{B. R. Ortiz}
\affiliation{Materials Science and Technology Division, Oak Ridge National Laboratory, Oak Ridge, TN 37831, USA}

\author{S. D. Wilson}
\affiliation{Materials Department and California Nanosystems Institute, University of California Santa Barbara, Santa Barbara, CA, 93106, USA}

\author{M. Dressel}
\affiliation{1. Physikalisches Institut, Universit{\"a}t Stuttgart, 70569 
Stuttgart, Germany}
\date{\today}

\begin{abstract}
Shifting the Fermi level of the celebrated $AM_3X_5$ (135) compounds into proximity of flat bands strongly enhances electronic correlations and severely affects the formation of density waves and superconductivity. Our broadband infrared spectroscopy measurements of \RTB\ and \CTB\ combined with density-functional band-structure calculations reveal that the correlated Ti $d$-states are intricately coupled with the Bi $p$-states that form a tilted Dirac crossing. Electron-phonon coupling manifests itself in the strong damping of itinerant carriers and in the anomalous shape of the phonon line in \RTB. An anomaly in these spectral features around 150~K can be paralleled to the onset of nematicity detected by low-temperature probes. Our findings show that the materials with low band filling open unexplored directions in the physics of kagome metals and involve electronic states of different nature strongly coupled with lattice dynamics.
\end{abstract}

\maketitle
Kagome 135-compounds have been the focal point of studying charge ordering, superconductivity, and electronic nematicity, among other exotic quantum phenomena in correlated electronic systems, over the last five years \cite{Wilson2024, Nguyen2022, Jiang2022, Neupert2022}. Rooted in the presence of flat bands, band saddle points (van Hove singularities), and Dirac bands, the rich electronic phase diagram of kagome metals is highly sensitive to the band filling level \cite{Denner2021, Kiesel2013, Park2021, Wang2013, Profe2024}. Building on the broad interest in the \AVS\ ($A$~=~K, Rb, Cs) series, new materials maintaining the pristine 135-kagome lattice structure have recently been proposed \cite{Wilson2024, XinWei2022, Si2024, Liu2024, Werhahn2022, Hu2024}. Among these, the \ATB\ ($A$~=~Rb, Cs) compounds stand out as one of the few synthesized materials with a significantly lower band filling level compared to \AVS\ compounds, according to the reduced number of valence electrons. Here, kagome-Ti planes stabilized by Bi1 atoms are stacked along the $c$-axis and separated by alkali ion layers and Bi2 honeycomb nets as sketched in Fig.~S1 \cite{SM}

Magnetic susceptibility, dc resistivity, and specific heat studies \cite{Werhahn2022, Wang2023} indicate the absence of a charge-density-wave (CDW) state in \ATB, a finding that is further corroborated by first-principle calculations \cite{Liu2023}. Some dc resistivity and magnetic susceptibility data report traces of a superconducting transition near 4~K \cite{Werhahn2022, Yang2024, Hu2023}. However, this observation remains contentious, as the transition temperature coincides with the $T_{\mathrm{c}}$ of possible RbBi$_2$ and CsBi$_2$ impurities \cite{Huan2022, Sankaralingam1992, Werhahn2022}. For \CTB, superconductivity emerges under external pressure, exhibiting a characteristic double-dome feature in $T{\mathrm{c}}$, similar to \AVS\ compounds, but with the second dome terminating at 35~GPa \cite{Wu2024, Nie2024}.

Scanning tunneling microscopy (STM) and angle-resolved photoemission spectroscopy (ARPES) studies reveal a reduction in the $C_6$ lattice symmetry to $C_2$ at low temperatures, without breaking the translational symmetry \cite{Yang2024, Li2023, Jiang2023, Hu2023}. This rotational symmetry breaking, without the formation of a CDW state, has been identified as a Pomeranchuk instability \cite{Bigi2024} and provides a path to explore purely electronic nematic order in nonmagnetic kagome metals. However, the exact temperature of this nematic transition and its occurrence in the bulk remain to be established, because the nematicity has been so far detected with surface-sensitive techniques and at low temperatures only.

In this work, we present a broadband optical spectroscopic study of \RTB\ and \CTB, probing the temperature-dependent evolution of the bulk electronic structure above and below the Fermi level. Aided by density functional theory (DFT) calculations, we demonstrate that the low-energy optical response is predominantly governed by interband transitions involving linearly dispersing Ti-kagome bands, with significant contributions from Bi2 $p_z$ states. Our findings highlight the enhancement of correlation effects in \ATB\ compared to \CVS\ and elucidate the role of Ti-flat bands in the vicinity of the Fermi level. Finally, we identify a change in the electron-phonon coupling in \RTB\ around 150~K and suggest that it may indicate the onset of nematicity.

\begin{figure*}
      \centering
      \includegraphics[width=2\columnwidth]{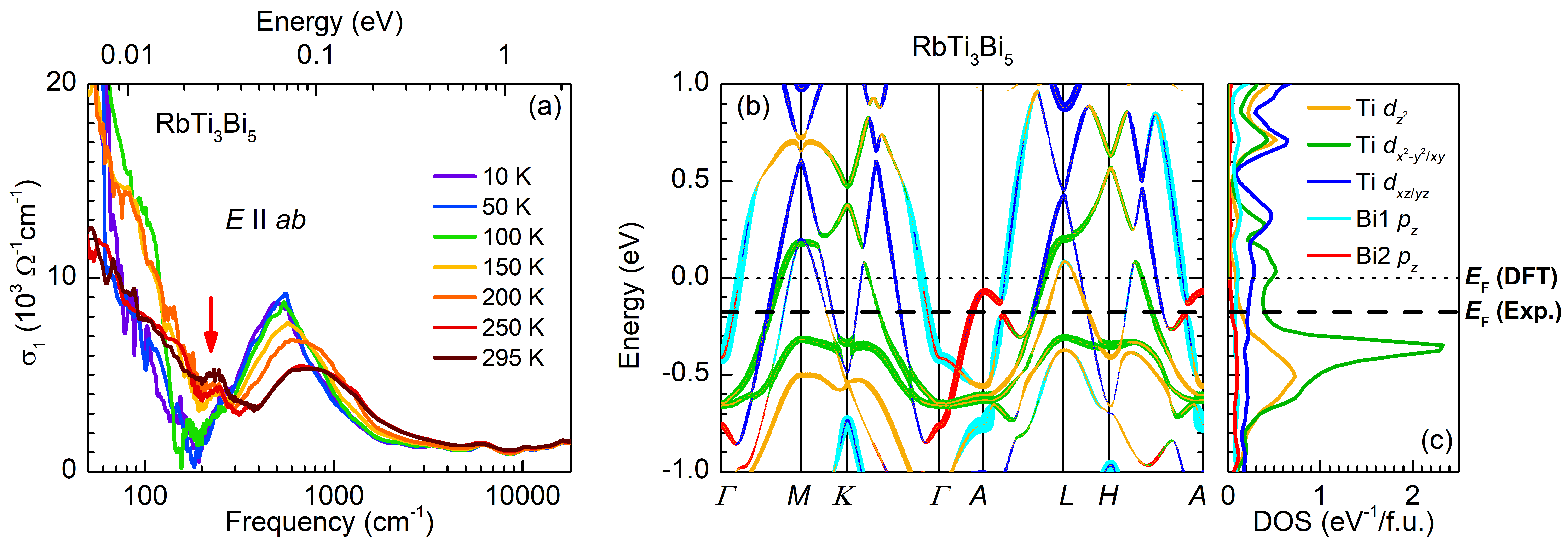}
  \caption{(a) Temperature-dependent real part of the in-plane optical conductivity of \RTB\ calculated from the measured reflectivity. The red arrow marks the Fano resonance around 200~\cm. (b) Band structure of \RTB\ with different colors representing contributions from different atomic orbitals as labeled in panel (c) showing the corresponding density of states (DOS). The short-dashed line shows the Fermi level obtained from DFT, whereas the long-dashed line is the position of the Fermi level adjusted using the experimental optical conductivity. }	
    \label{Fig1}
\end{figure*}

High-quality single crystals were synthesized using a self-flux method \cite{Werhahn2022}. Reflectivity measurements were performed in the $ab$-plane on as-grown single crystals in a broad frequency range (50~-~18000~\cm) at different temperatures down to 10~K. The optical conductivity was obtained by Kramers-Kronig analysis as explained in the Supplemental Material \cite{SM}. Fig.~\ref{Fig1}(a) shows the calculated real part of the optical conductivity of \RTB\ at selected temperatures. The raw data as well as the spectra of \CTB\ can be found in the Supplemental Material \cite{SM}. 

Band structure and optical conductivity calculations were performed in the \texttt{Wien2K} code \cite{wien2k, Blaha2020} using the experimental crystal structures from Ref.~\cite{Werhahn2022} and the Perdew-Burke-Ernzerhof (PBE) exchange-correlation potential \cite{pbe96}. Spin-orbit coupling was included in all calculations. The optical conductivity was obtained using the \texttt{OPTIC} module of \texttt{Wien2K} \cite{Draxl2006}.

Fig.~\ref{Fig1}(b) displays the calculated band structure of \RTB\ highlighting the different orbital contributions. The low-energy band structure is dominated by linearly dispersing Ti $d$- and Bi1 $p_z$-states crossing the Fermi level. Band saddle points, comprised mainly of Ti $d_{x^2-y^2/xy}$ states, are located above the Fermi level at the $M$- and $L$-points. Additionally, flat bands with a large density of Ti $d_{x^2-y^2/xy}$ and $d_{z^2}$ states [see panel (b)] are present below the Fermi level.

\begin{figure}
\includegraphics[width=1\columnwidth]{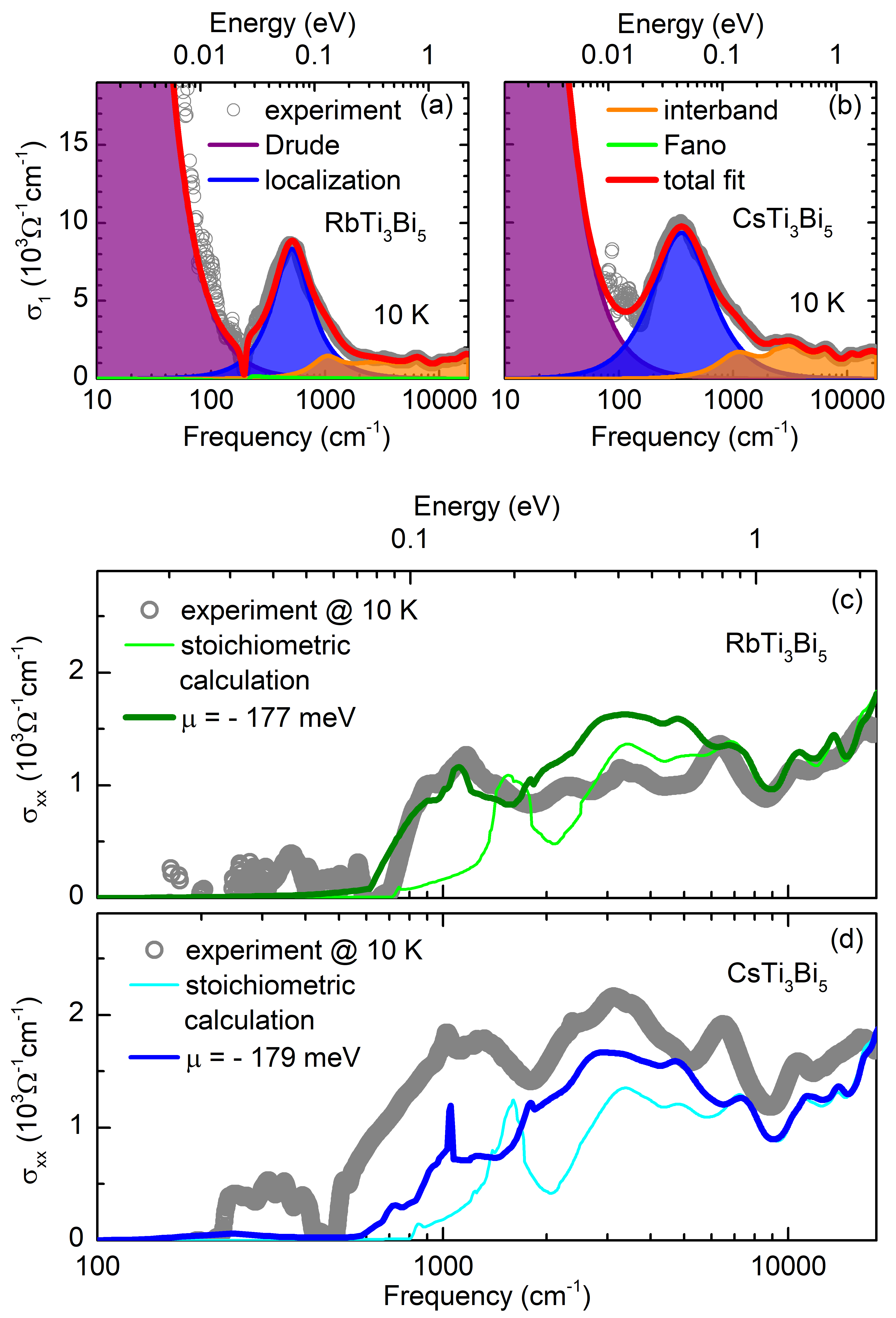}
\caption{Comparison of the optical response of \RTB\ and \CTB. (a) and (b) Decomposed optical conductivities at 10~K, consisting of a Drude peak (purple), a localization peak (blue), several interband transitions (orange), and a Fano resonance (green) in the case of \RTB. (c) and (d) Experimental and DFT-based in-plane interband optical conductivities of \RTB\ and \CTB, respectively. Above 2500~\cm, a broadening of 0.05~eV is applied to the calculated spectra. The shift of the Fermi level necessary to reproduce the experimental results is expressed by the chemical potential $\mu$. }
    \label{Fig2}
\end{figure}

Experimental optical conductivities are analyzed by modeling the different contributions to the total optical spectra using the Drude-Lorentz approach as presented in Figs.~\ref{Fig2}(a) and (b) for the 10~K data. In addition to the classical Lorentzian (interband) and Drude (intraband) contributions, a strong, temperature-dependent absorption peak is observed at low energies (blue), which is attributed to the intraband signature of localized electrons \cite{Wenzel2022a, Uykur2021, Uykur2022, Fratini2014, Fratini2021}. The spectra of \RTB\ further reveal a rather broad Fano-shaped phonon mode around 200~\cm, marked by the red arrow in Fig.~\ref{Fig1}(a), signaling the presence of strong electron-phonon coupling. The total complex optical conductivity [$\tilde{\sigma}=\sigma_1 + i\sigma_2$] then takes the form
\begin{equation}
\tilde{\sigma}(\omega)= \tilde{\sigma}_{\mathrm{intraband}} + \tilde{\sigma}_{\mathrm{phonon}} + \tilde{\sigma}_{\mathrm{interband}}.
\label{Cond}
\end{equation}
The experimental interband optical transitions are obtained by subtracting the intraband contributions, i.e., Drude and localization peaks, and the Fano resonance from the optical conductivity. Figs.~\ref{Fig2}(c) and (d) show the experimental and DFT-based interband optical conductivity of \RTB\ and \CTB, respectively. Stoichiometric calculations predict a sharp absorption feature around 1500~\cm, which deviates from the experimental data, indicating the need for band energy renormalization. A better agreement between experiment and theory is achieved by a downward shift of the Fermi level by 177~meV for \RTB\ and 179~meV for \CTB. This adjustment reproduces the experimental low-energy absorption observed near 1000~\cm. Notably, a similar downward shift of the Fermi level has also been reported in ARPES measurements for the Cs compound \cite{Yang2023}.

\begin{figure}
        \centering
       \includegraphics[width=1\columnwidth]{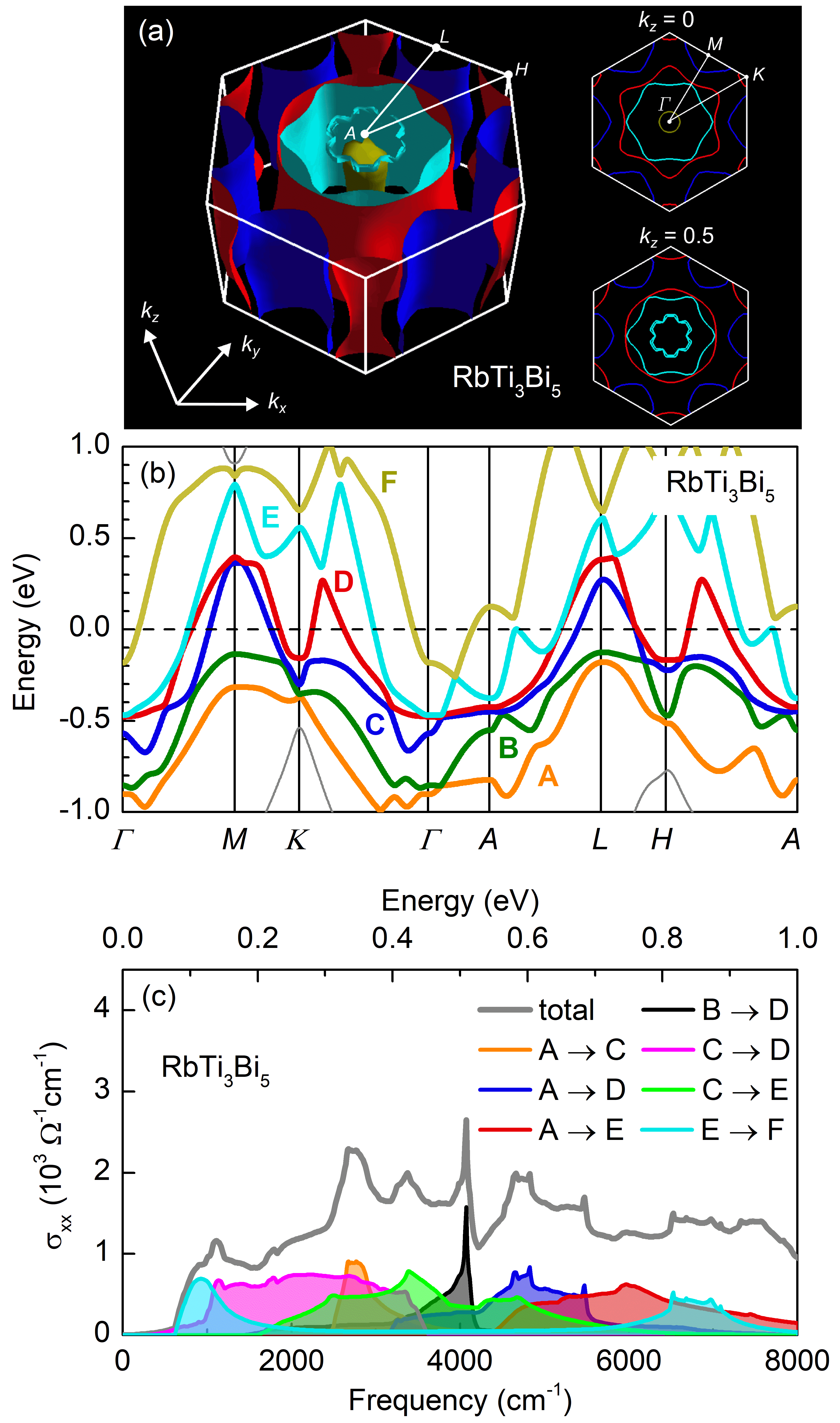}
   \caption{(a) Fermi surface of \RTB\ constructed from the calculated band structure with the Fermi level shifted down by 177~meV to match the experimental data. \texttt{FermiSurfer} program was used for the visualization \cite{Kawamura2019}. (b) Calculated band structure of \RTB\ with the experimentally determined position of the Fermi level. (c) Band-resolved in-plane optical conductivity with different colors representing contributions from interband optical transitions between different bands following the labeling in panel (b). Note that no broadening is introduced to the calculations. }	
    \label{Fig3}
\end{figure}

Interestingly, this substantial downward shift of the Fermi level has minuscule impact on the bands crossing $E_{\mathrm{F}}$ along $\Gamma$-$M$-$K$-$\Gamma$, as outlined in Fig.~\ref{Fig1}(b). Therefore, the Fermi surface at $k_z = 0$ [see Fig.~\ref{Fig3}(a)] is largely unchanged and consistent with those obtained from ARPES and quantum oscillation studies \cite{Yang2023, Jiang2023, Dong2024, Rehfuss2024}. The primary effect of the Fermi level shift is the emergence of Bi2 $p_z$ states at the Fermi level around the $A$ point, as illustrated in Fig.~\ref{Fig1}(a). This leads to the formation of two new Fermi surface sheets around the $A$ point, depicted in Fig.~\ref{Fig3}(a) for \RTB.

To further investigate the contributions of different electronic bands to the interband conductivity, we perform band-resolved calculations, presented in Fig.~\ref{Fig3}(c) for \RTB. The interband absorption at the lowest energies arises purely from (i) transitions between the linearly dispersing Ti $d$ bands C and D, as labeled in Fig.~\ref{Fig3}(b), and (ii) transitions between the bands E and F. The latter transitions resemble a sharp Pauli edge and can be attributed to the tilted Dirac-like bands along $A$-$L$ and $H$-$A$, predominantly involving Bi1 and Bi2 $p_z$ orbitals [see Figs.~\ref{Fig1}(b) and (c)].Previous studies highlighted such tilted linear crossings along $\Gamma$-$M$ and $A$-$L$ where they comprise $d$-bands only (bands D and E in our notation) \cite{Yang2023, Jiang2023}. Our band-resolved calculations, however, reveal that these type-II Dirac crossings of the bands D and E do not contribute to the low-energy optical response, which is dominated by the E~$\rightarrow$~F transitions. The replacement of the alkali ion does not change this scenario (see Fig.~S9 \cite{SM}) but essentially only shifts the tilted Dirac bands of Bi relative to the Fermi energy as highlighted in Fig.~S8 \cite{SM}. Consequently, band E forms hole pockets in the Fermi surface of \RTB\ at $k_z = 0.5$, while the respective pockets in \CTB\ are instead created by band F and exhibit electron character.

The large shift of the Fermi level required to reproduce the experimental optical conductivity moves the Ti flat bands closer to the Fermi level, positioning them at approximately \mbox{-0.2~eV} and \mbox{-0.4~eV}, in good agreement with ARPES measurements \cite{Yang2023, Jiang2023}. The flat bands are even closer to the Fermi level in CsCr$_3$Sb$_5$ where the Cr $d$-bands exhibit strong orbital-selective renormalization effects \cite{Xie2024, Guo2024, Li2024}. The treatment of correlation effects in \ATB\ within the DFT+$U$ framework is discussed in the Supplemental Material \cite{SM}. However, adding a Hubbard $U$ term to the Ti-$d$ orbitals pushes the flat bands farther away from the Fermi energy and induces modifications in the Fermi surface, failing to reproduce both the electronic band structure observed in ARPES studies and the optical response. This suggests more complex, orbital-selective correlation effects similar to the ones observed in CsCr$_3$Sb$_5$ that cannot be captured within the mean-field DFT+$U$ approach.

\begin{figure}
        \centering
       \includegraphics[width=1\columnwidth]{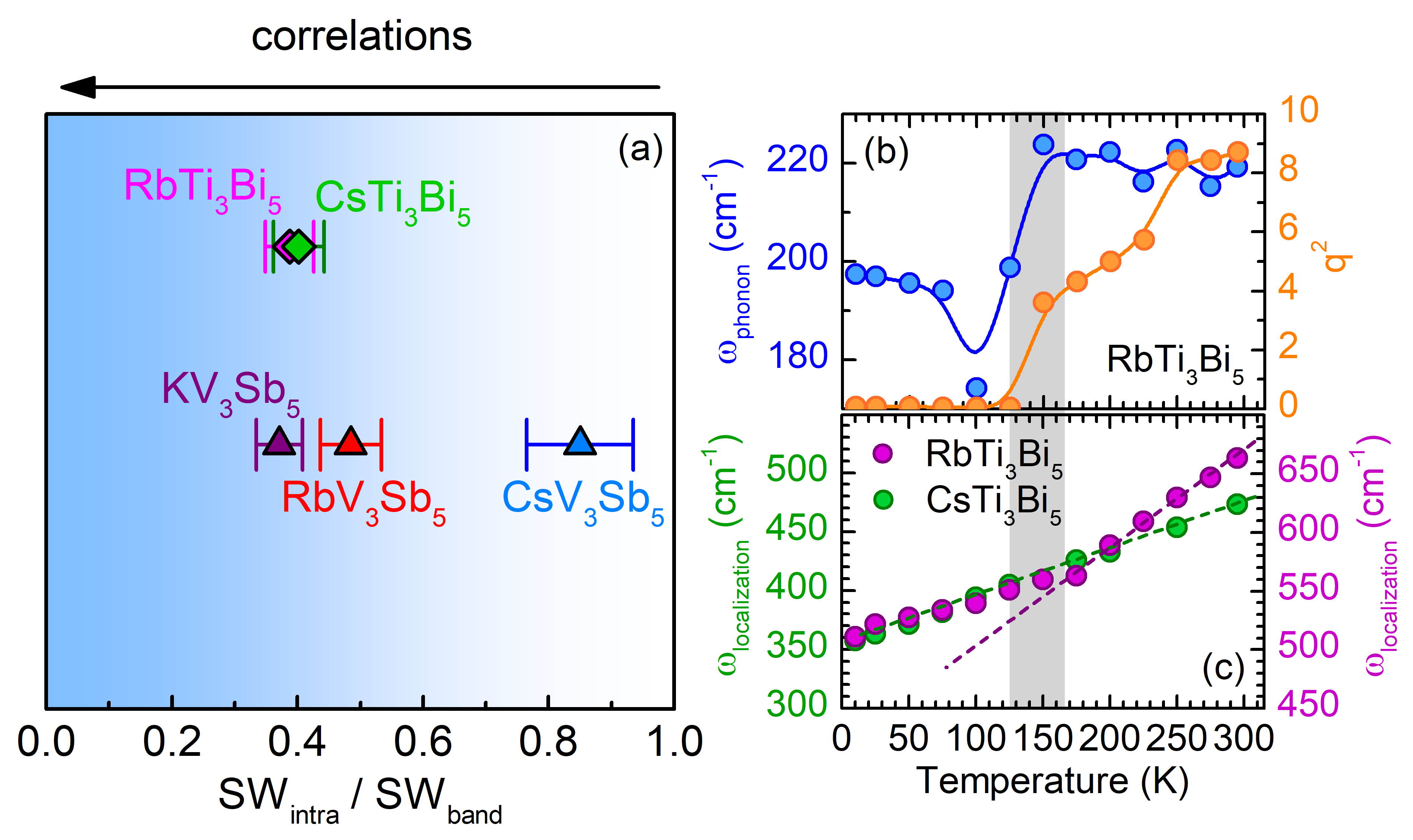}
   \caption{(a) Ratio of the experimental and DFT-based intraband spectral weights serving as a gauge of the strength of electronic correlations. Values for the vanadium-based 135-compounds are taken from previous studies \cite{Uykur2022, Uykur2021, Wenzel2022a}. Error bars of 10~\% arising from uncertainties in determining the experimental plasma frequency are assumed as elaborated in the Supplemental Material \cite{SM}. (b) Temperature evolution of the phonon frequency observed in the optical spectra of \RTB\ (blue) and the Fano coupling parameter $q^2$ (orange). (c) Comparison of the localization peak position in \RTB\ (right scale) and \CTB\ (left scale) as a function of temperature. Two distinct slopes are highlighted by dashed lines.}
    \label{Fig4}
\end{figure}

Significant electronic correlations in \ATB\ are clearly witnessed by our optical data. Not only must the band energy be renormalized by the downward shift of the Fermi level to align with the experimental interband transitions, but also the calculated plasma frequencies of 4.02~eV and 3.98~eV for \RTB\ and \CTB, respectively, are much higher than the experimental values of 2.50~eV (Rb) and 2.52~eV (Cs). This discrepancy indicates that DFT overestimates the spectral weight of itinerant charge carriers. The strength of the electronic correlations can be gauged by the ratio of the experimental to DFT-based intraband spectral weights \cite{Shao2020, Qazilbash2009}, as outlined in the Supplemental Material \cite{SM}. This ratio is approximately 0.40 for both \RTB\ and \CTB\ in stark contrast to \CVS, which is almost uncorrelated as illustrated in Fig.~\ref{Fig4}(a). Interestingly, though, \KVS\ and \RVS\ also show enhanced correlations.

The experimental optical spectra of \RTB\ reveal a relatively broad phonon mode around 200~\cm, displaying Fano-like behavior, a hallmark of electron-phonon coupling. Such phonon modes are relatively common in kagome metals and have been observed, albeit at higher energies, in other 135-compounds as well \cite{Wenzel2022a, Uykur2022}. According to our DFT calculations outlined in the Supplemental Material \cite{SM}, the mode in \RTB\ can be assigned to the infrared-active $E_{\mathrm{1}u}$ phonon mainly involving in-plane atomic displacements within the Ti-Bi1 kagome net. 

Below 150~K, the phonon spectrum evolves into a pronounced anti-resonance seen in Fig.~S6 \cite{SM}, which corresponds to a drastic change in the Fano coupling parameter $q^2$ plotted in Fig.~\ref{Fig4}(b) (orange). Simultaneously, a sudden shift in the phonon frequency to lower energies is observed [Fig.~\ref{Fig4}(b) (green)]. While changes in phonon spectra are often associated with structural phase transitions, no alterations in the interband optical absorption are observed across the entire temperature range [see Fig.~S4 for the decomposed spectra at different temperatures \cite{SM}]. Combined with the absence of a low-temperature anomaly in dc transport, magnetic susceptibility, or specific heat \cite{Werhahn2022, Wang2023}, these findings make structural changes unlikely. Additionally, we observe an anomaly in the position of the localization peak around the same temperature of 150~K [Fig.~\ref{Fig4}(c)]. The localization peaks are typically caused by electrons interacting with bosonic degrees of freedom, such as phonons \cite{Rammal2024, Biswas2020, Wenzel2023}. Therefore, we interpret the changes in the phonon line and the anomaly in the position of the localization peak as a sign of sudden changes in the electron-phonon coupling upon entering the nematic phase at low temperatures. Consequently, our data suggest a possible onset of bulk nematicity in \RTB\ just below 150~K. The absence of comparable signatures in \CTB\ does not necessarily rule out a similar scenario in this material. Rather, it may result from the phonon mode being screened by conduction electrons. However, the lack of an anomaly in the localization peak position in \CTB\ could indicate that nematicity in this system is confined to the surface. On the other hand, the temperature dependence of the localization peak in \CTB\ exhibits a slope strikingly similar to that of \RTB\ in its low-temperature (nematic) phase [see Fig.~4(c)], possibly suggesting the onset of nematicity already above room temperature in \CTB. 

Overall, the significantly reduced band filling level in the \ATB\ compounds compared to their \AVS\ counterparts strongly modifies the nature of low-energy optical excitations. In the case of the \AVS, the low-energy electrodynamics are dominated by charge carriers around the $M$ point, which arise from vanadium 3$d$ kagome states \cite{Wenzel2022a, Uykur2021, Uykur2022, Tsirlin2022, Mozaffari2024, Peshcherenko2024}. In contrast, non-kagome Bi $p_z$ states forming a tilted Dirac crossing play a substantial role in \ATB. These states are witnessed by the low-energy spectral weight observed in our measurements. Additionally, the presence of flat bands can be inferred from the significant band renormalization and the reduced intraband spectral weight, both indicative of the sizable electronic correlations in \ATB.

The proximity of the Fermi level to the flat bands seems crucial for the \ATB\ physics that entails orbital-selective correlations leading to the $d$-wave Pomeranchuk instability and a purely electronic nematic phase \cite{Bigi2024}. Our results extend this picture in two aspects. First, we demonstrate the presence of electron-phonon coupling evidenced by the anomalous phonon mode and the prominent localization peak of the intraband carriers. Second, we show that both of these features reveal an anomaly around 150~K, which is likely associated with the nematic transition. Our observations suggest that phonons can serve as useful fingerprints of electronic instabilities and call for momentum-dependent phonon studies on \ATB.

Our findings underscore the prominent role of linear and flat Ti 3$d$ as well as Bi $p_z$ states in shaping the low-energy electronic structure of the \ATB\ compounds. Especially the presence of Bi2 $p_z$ Fermi surface pockets clearly distinguishes these compounds from their \AVS\ counterparts. Simultaneously, the effects of electronic correlations are amplified due to the proximity of Ti flat bands to the Fermi level. The temperature evolution of the optical spectra reveals an anomaly around 150~K that may be associated with the onset of nematicity.
\FloatBarrier

\textit{Note added.} During the completion of this work, an independent infrared spectroscopy study \cite{Cao2024} reported optical signatures of flat bands in \CTB\ similar to our work, but neither the low-energy features presented here nor the correlation strength could be resolved in that study.

We are grateful to Gabriele Untereiner for preparing single crystals and assisting in the optical measurements. M.W. is supported by Center for Integrated Quantum Science and Technology (IQST) Stuttgart/Ulm via a project funded by the Carl Zeiss Stiftung. Work by B.R.O. was supported by the U.S. Department of Energy, Office of Science, Basic Energy Sciences, Materials Sciences and Engineering Division. S.D.W gratefully acknowledges support via the UC Santa Barbara NSF Quantum Foundry funded via the Q-AMASE-i program under award DMR-1906325. The work has been supported by the Deutsche Forschungsgemeinschaft (DFG) via DR228/51-3, DR228/68-1, and UY63/2-1.

\end{document}


\title{Supplemental Material for "Interplay of $d$- and $p$-states in \RTB\ and \CTB\ flat-band kagome metals"}

\author{M. Wenzel}
\email{maxim.wenzel@pi1.physik.uni-stuttgart.de}
\affiliation{1. Physikalisches Institut, Universit{\"a}t Stuttgart, 70569 
Stuttgart, Germany}

\author{E. Uykur}
\affiliation{Helmholtz-Zentrum Dresden-Rossendorf, Institute of Ion Beam Physics and Materials Research, 01328 Dresden, Germany}

\author{A. A. Tsirlin}
\affiliation{Felix Bloch Institute for Solid-State Physics, Leipzig University, 04103 Leipzig, Germany}

\author{A. N. Capa Salinas}
\affiliation{Materials Department and California Nanosystems Institute, University of California Santa Barbara, Santa Barbara, CA, 93106, USA}

\author{B. R. Ortiz}
\affiliation{Materials Science and Technology Division, Oak Ridge National Laboratory, Oak Ridge, TN 37831, USA}

\author{S. D. Wilson}
\affiliation{Materials Department and California Nanosystems Institute, University of California Santa Barbara, Santa Barbara, CA, 93106, USA}

\author{M. Dressel}
\affiliation{1. Physikalisches Institut, Universit{\"a}t Stuttgart, 70569 
Stuttgart, Germany}
\date{\today}

\setcounter{equation}{0}
\setcounter{figure}{0}
\setcounter{table}{0}
\setcounter{page}{1}
\makeatletter
\renewcommand{\theequation}{S\arabic{equation}}
\renewcommand{\thefigure}{S\arabic{figure}}
\renewcommand{\bibnumfmt}[1]{[S#1]}
\renewcommand{\citenumfont}[1]{S#1}
\maketitle
\section{Crystal Structure}
\begin{figure}[h]
\includegraphics[width=1\columnwidth]{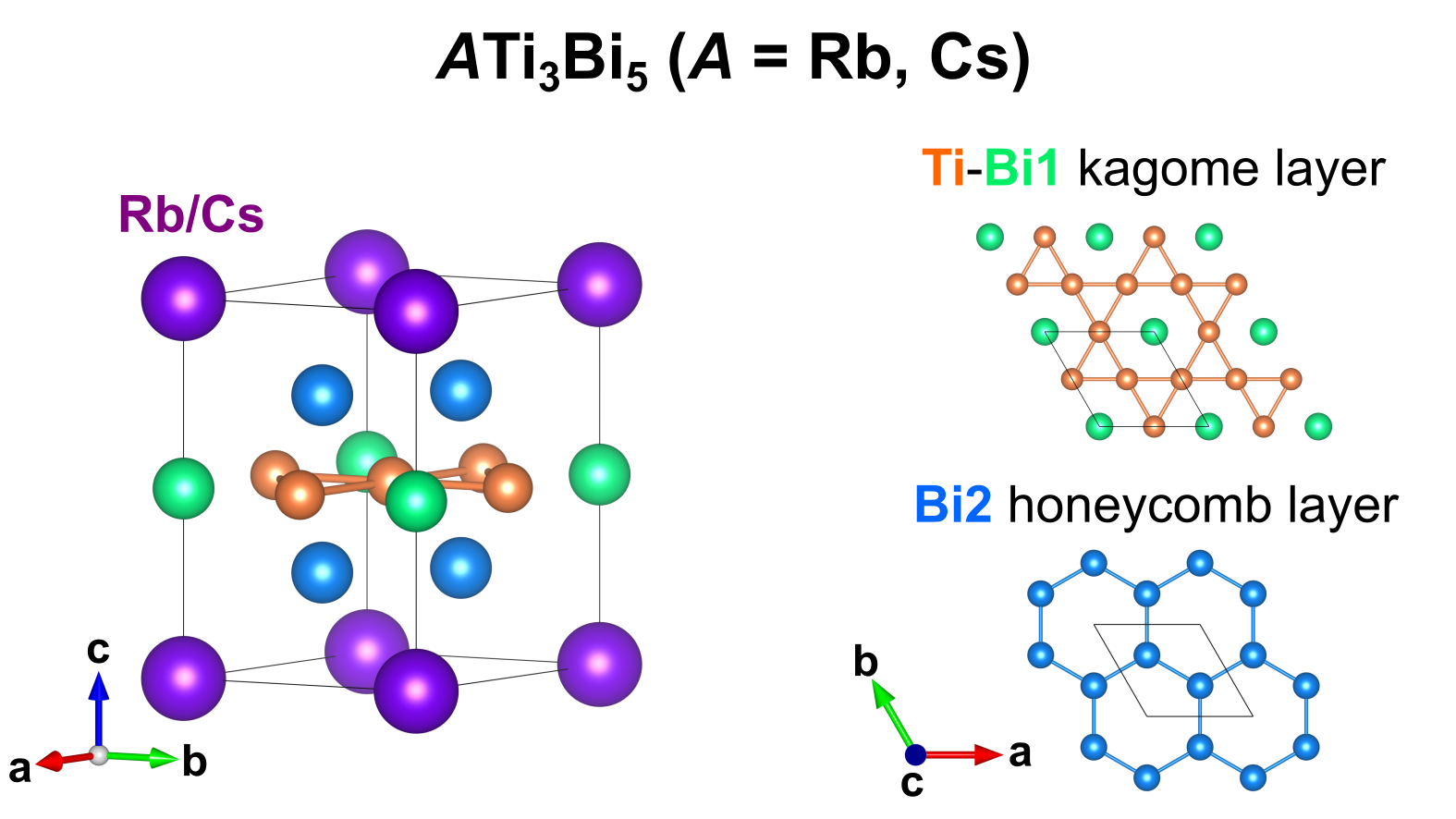}
\caption{Crystal structure of \ATB\ compounds visualized using \texttt{VESTA} \cite{Momma2008, Werhahn2022}. The right panel highlights the two different Bi sublattices.}
    \label{structure}
\end{figure}

\ATB\ compounds crystallize in the \textit{P6/mmm} space group with Ti atoms forming  kagome layers stacked along the $c$-axis. Bi1 atoms occupy the centers of the hexagons within these kagome networks, while a second Bi2 honeycomb sublattice, along with alkali ions, spatially separates the kagome layers. This arrangement results in well-isolated kagome planes and an overall quasi-two-dimensional (2D) structure, as shown in Fig.~\ref{structure}.

\section{Experimental details}
\begin{figure}
\includegraphics[width=1\columnwidth]{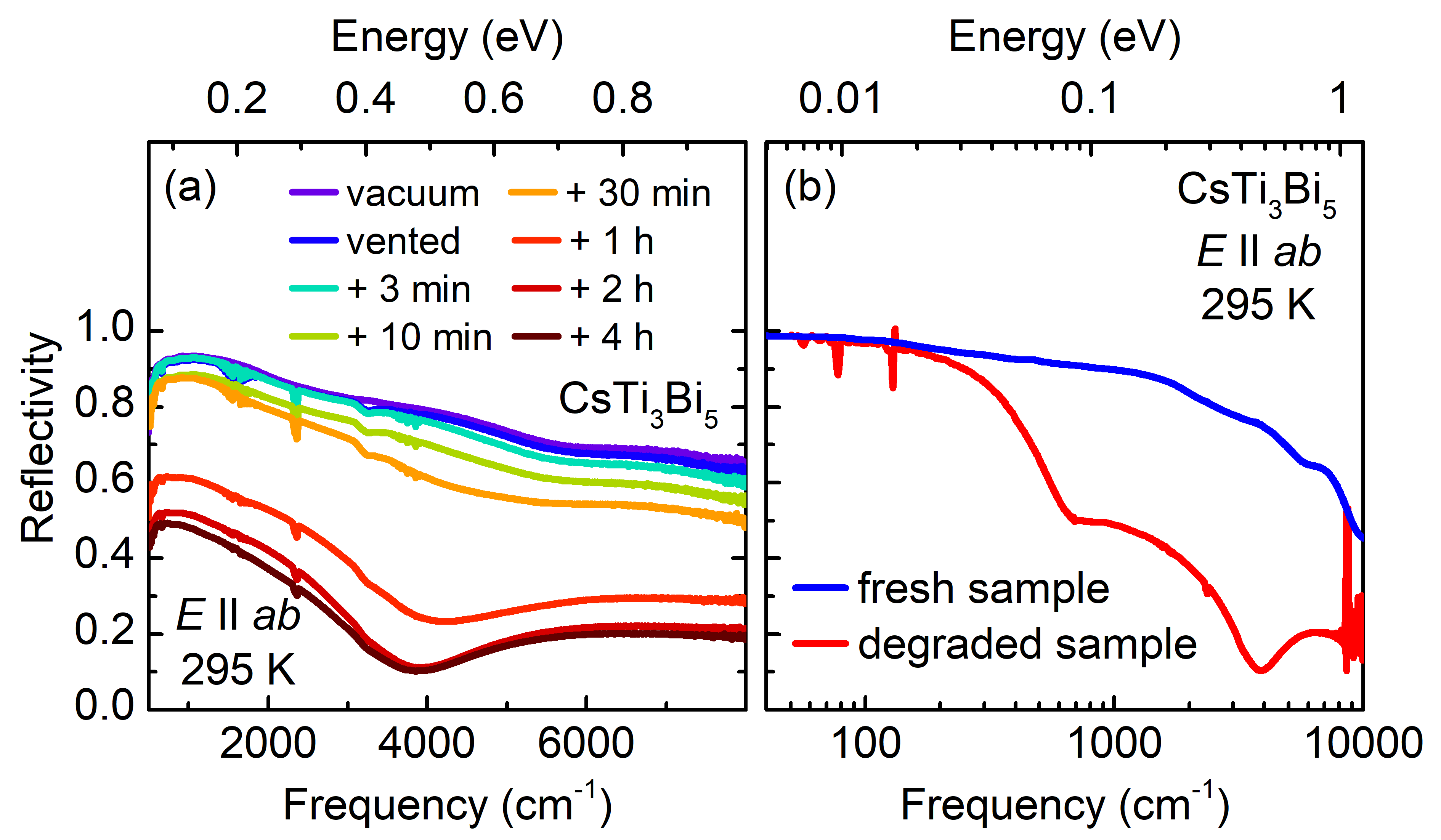}
\caption{(a) Reflectivity of \CTB\ in the mid-infrared range measured as a function of exposure time of the sample to air. The absolute curves are normalized to the far-infrared spectra. (b) Far- to mid-infrared spectra of \CTB\ for a fresh samples and a degraded sample. For the later case the sample was exposed to air for $>$~4~h. }
    \label{FigS1}
\end{figure}

Samples with lateral dimensions of $\sim$~2~$\times$~3~mm$^2$ were freshly cleaved and prepared in a glove box under argon atmosphere. Prior to conducting temperature-dependent reflectivity measurements, we tested the air sensitivity of \CTB, summarized in Fig.~\ref{FigS1}(a). Here, the sample was exposed to air, and mid-infrared spectra were collected over a four-hour span. Within just three minutes, a noticeable dip began to form around 4000~\cm, which continued to increase upon longer exposure, indicating substantial alterations in the bulk electronic structure of the material. The far-infrared reflectivity spectrum of a degraded \CTB\ sample, collected four hours post-exposure to air, is presented in panel~(b), merged with the mid-infrared data. The combined data reveal a substantial decrease in reflectivity already around 300~\cm, suggesting a pronounced shift of the plasma frequency to lower energies and, consequently, a significant reduction in carrier density. Visually, this degradation process leads to tarnishing of the sample's surface within just a few hours.

\begin{figure*}
\includegraphics[width=1.75\columnwidth]{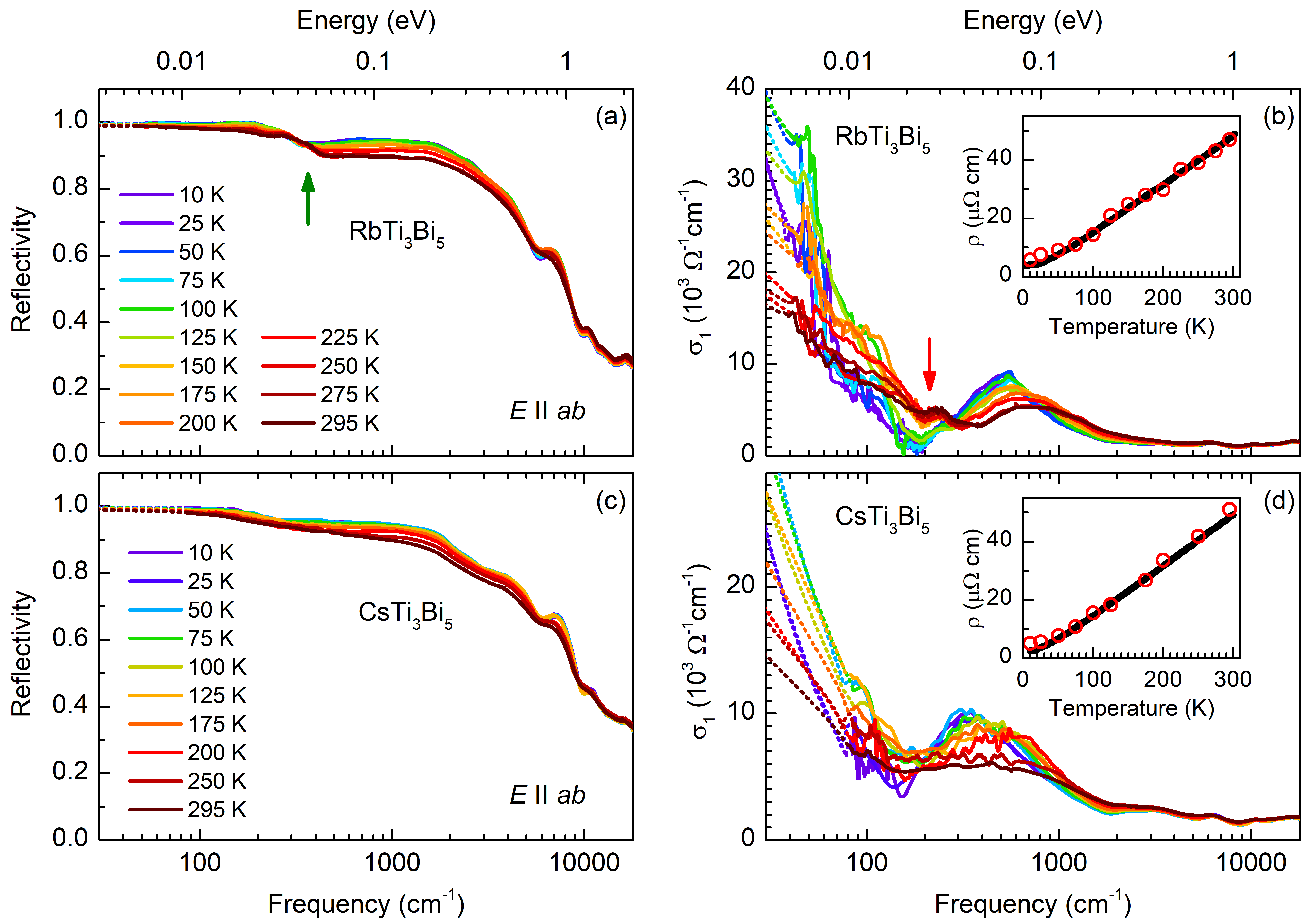}
\caption{(a) and (c) Temperature-dependent reflectivities of \RTB\ and \CTB, respectively, measured in the $ab$-plane. Dotted lines indicate the Hagen-Rubens extrapolations to low energies. The green arrow marks the sudden drop in reflectivity leading to the sharp onset of the localization peak. (b) and (d) Calculated real part of the optical conductivities. The red arrow indicates the Fano resonance observed in the spectra of \RTB. The insets show the dc resistivity taken from Ref.~\cite{Werhahn2022} overlapped with the dc resistivity values obtained from the Hagen-Rubens fits of the measured spectra (open red circles). The dc resistivity curve was normalized to the values obtained by the optical studies.}
    \label{FigS2}
\end{figure*}

Temperature- and frequency-dependent reflectivity spectra were measured in the $ab$-plane utilizing two different spectrometer setups. For high energies ($600\, \mathrm{cm^{-1}} \leq \omega \leq 18000$~\cm), a Vertex80v spectrometer coupled to a Hyperion IR microscope was used. Freshly evaporated gold mirrors served as the reference throughout the measurements. The low-energy range ($50\, \mathrm{cm^{-1}} \leq \omega < 600$~\cm) was measured with an IFS113v spectrometer and a custom-built cryostat. To prevent degradation, the samples were mounted into the cryostats and aligned within the glovebox. The cryostat used for measurements in the high-energy range was sealed with aluminum tape, while a custom-built sleeve was used for the cryostat designed for far-infrared measurements, maintaining an argon atmosphere inside. The cryostats were then transferred to the experimental setups and promptly evacuated, reaching a pressure of $p < 10^{-2}$~mbar in under two minutes.

The absolute value of the reflectivity was determined with higher accuracy using the gold-overcoating technique as described in Ref.~\cite{Homes1993} in the far-infrared range. However, using the usual gold evaporation parameters resulted in a tarnished gold coating, indicating a chemical reaction. Hence, for measurements in the far-infrared range, we chose to evaporate aluminum that preserved the shiny metallic surface of the samples, and used freshly prepared aluminum mirrors as reference.

Figs.~\ref{FigS2}(a) and (c) present the temperature-dependent reflectivity spectra of \RTB\ and \CTB, respectively. The high reflectivity values throughout all temperatures demonstrate the metallic nature of the samples and assure that the samples remained intact without any signs of oxidation or degradation throughout the measurements. The Hagen-Rubens extrapolations at the low frequency limit are plotted with dotted lines. The obtained dc resistivity values match well with the four-probe dc resistivity adapted from Ref.~\cite{Werhahn2022} after normalization of the data [see insets of Figs.~\ref{FigS2}(b) and (d)]. For high-energies, the reflectivity data are extrapolated using x-ray scattering functions \cite{Tanner2015}. The optical conductivities calculated via standard Kramers-Kronig analysis \cite{Dressel2002} are given in Figs.~\ref{FigS2}(b) and (d). 

\section{Decomposition}
Different contributions to the total optical conductivity are modeled with the Drude-Lorentz approach. The dielectric function [$\tilde{\varepsilon}=\varepsilon_1 + i\varepsilon_2$] is expressed as
\begin{equation}
\label{Eps}
\tilde{\varepsilon}(\omega)= \varepsilon_\infty - \frac{\omega^2_{p,{\rm Drude}}}{\omega^2 + i\omega/\tau_{\rm\, Drude}} + \sum\limits_j\frac{\Omega_j^2}{\omega_{0,j}^2 - \omega^2-i\omega\gamma_j},
\end{equation}
with $\varepsilon_{\infty}$ being the high-energy contributions to the real part of the dielectric permittivity. The Drude parameters $\omega_{p,{\rm Drude}}$ and $1/\tau_{\rm\,Drude}$ describe the plasma frequency and the scattering rate of the itinerant carriers, respectively. Lorentzians with the resonance frequency $\omega_{0,j}$, the strength of the oscillation $\Omega_j$, and the width $\gamma_j$ are used to model the interband absorptions.

The complex optical conductivity [$\tilde{\sigma}=\sigma_1 + i\sigma_2$] is calculated as
\begin{equation}
\tilde{\sigma}(\omega)= -i\omega \varepsilon_{\mathrm{0}}[\tilde{\varepsilon} (\omega) -\varepsilon_\infty],
\end{equation}
with $\varepsilon_{\mathrm{0}}$ being the vacuum permittivity.

In addition to the classical Drude-Lorentz approach, we use the displaced Drude peak formalism proposed in 2014 by Fratini et \textit{al.} \cite{Fratini2014} to model the intraband response of localized charge carriers. Here, interactions of charge carriers with low-energy degrees of freedom, such as phonons and electric or magnetic fluctuations, can lead to a backscattering of the electrons, with a characteristic time $\tau_ {\mathrm{b}}$. The resulting localization effect shifts the classical Drude peak from zero frequency to a finite value. The occurrence of such broad, low-energy, temperature-dependent absorption peaks is a common feature in strongly correlated electron systems \cite{Fratini2021} and is also characteristic for the optical spectra of kagome metals \cite{Biswas2020, Uykur2021, Uykur2022, Wenzel2022, Wenzel2022a}. The real part of the optical conductivity reads
\begin{multline}
\label{Fratini}
\tilde{\sigma}_{\rm localization}(\omega)=\frac{C}{\tau_{\mathrm{b}}-\tau}\frac{\tanh\{\frac{\hbar\omega}{2k_{\mathrm{B}}T}\}}{\hbar\omega}\\ \cdot\,\mathrm{Re}\left\{\frac{1}{1-\mathrm{i}\omega\tau}-\frac{1}{1-\mathrm{i}\omega\tau_{\mathrm{b}}}\right\},
\end{multline}
where $C$ is a constant, $\hbar$ is the reduced Planck constant, $k_{\mathrm{B}}$ the Boltzmann constant, and $\tau$ the elastic scattering time of the standard Drude model. 
\begin{figure}
\includegraphics[width=1\columnwidth]{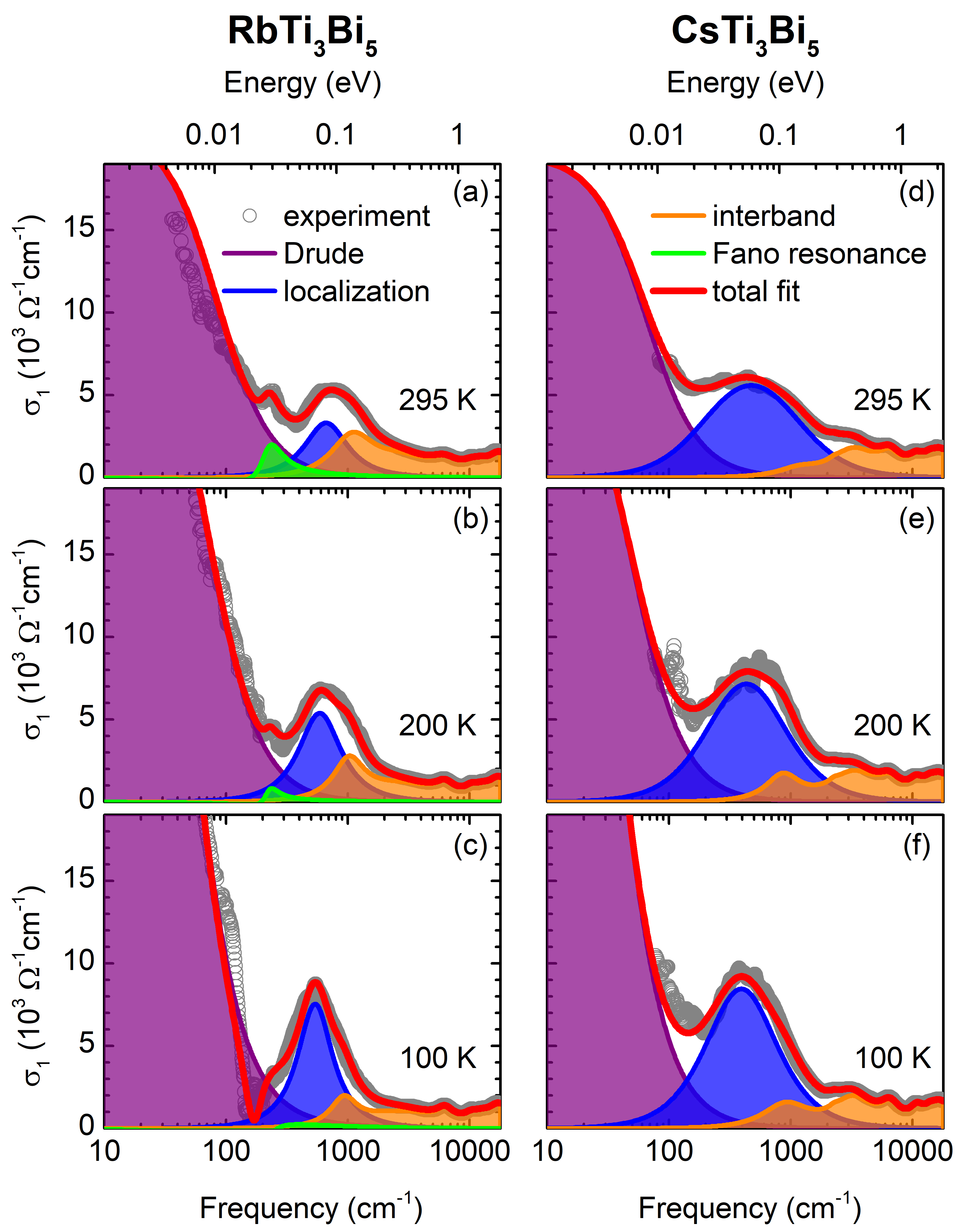}
\caption{Decomposed optical conductivities of \RTB\ (a-c) and \CTB\ (d-f) at selected temperatures, consisting of a Drude peak (purple), a localization peak (blue), several interband transitions (orange), and a Fano resonance (green) in the case of \RTB. }
    \label{FigS3}
\end{figure} 

The asymmetric phonon mode observed in the spectra of \RTB, at approximately 200~\cm, is reproduced most accurately by the Fano model \cite{Fano1961}
\begin{equation}
\sigma_1(\omega) = \frac{\sigma_0\omega\gamma[\gamma\omega (q^2-1)+2q(\omega^2-\omega_0^2)]}{4\pi[(\omega^2-\omega_0^2)^2+\gamma^2\omega^2]} \quad .
 \label{fano}
\end{equation}
Here, $\sigma_{\mathrm{0}}$ is the resonance strength, $\gamma$ the damping and $\omega_{\mathrm{0}}$ the resonance frequency. The shape of the Fano resonance highly depends on the dimensionless parameter, $q$. The typical asymmetric line shape is obtained for moderate values of $|q|$, while for $|q| \rightarrow \infty$, the line evolves into a symmetric Lorentzian. For $q = 0$, a strong symmetric anti-resonance is observed.

The total optical conductivity takes the form
\begin{equation}
\tilde{\sigma}(\omega)= \underbrace{\tilde{\sigma}_{\mathrm{Drude}} + \tilde{\sigma}_{\mathrm{localization}}}_{\mathrm{intraband}} + \underbrace{\tilde{\sigma}_{\mathrm{Fano}}}_{\mathrm{phonon}} + \underbrace{\tilde{\sigma}_{\mathrm{Lorentzians}}}_{\mathrm{interband}}.
\end{equation}
The reliability of the fits is ensured by fitting the reflectivity, and the complex optical conductivity simultaneously. Fig.~\ref{FigS3} displays the decomposed real part of the optical conductivity at selected temperatures.

\section{Localization peak}
In the absence of low-energy interband transitions, the absorption peaks below 1000~\cm should be interpreted as localization peaks, similar to GdMn$_6$Sn$_6$ and \KVS\ \cite{Wenzel2022, Uykur2022}. When focusing solely on the room temperature data, clear differences between the localization peaks in \RTB\ and \CTB\ become apparent. In \RTB, the response from localized carriers is significantly less intense, albeit much sharper than in \CTB. These characteristics mirror those of the vanadium-based \AVS\ counter parts \cite{Wenzel2022a, Uykur2021}, suggesting that the alkali ions may influence the behavior of localized carriers, potentially due to interactions between the alkali ions and the kagome layers, facilitated by the large ionic radius of the alkali ions. 

In \RTB, the localization peak is so sharp that it cannot be modeled by Eq.~\ref{Fratini}. Specifically, the sharp onset of the peak at lower energies, which results from the sudden drop in reflectivity marked by the green arrow in Fig.~\ref{FigS2}(a), cannot be adequately captured by this model. Consequently, a simple Lorentzian was employed to fit the localization peak in \RTB. Nonetheless, it is important to emphasize that these mathematical challenges in modeling the localization peak do not alter the fundamental understanding of its underlying physical mechanism. 

\begin{figure}
\includegraphics[width=1\columnwidth]{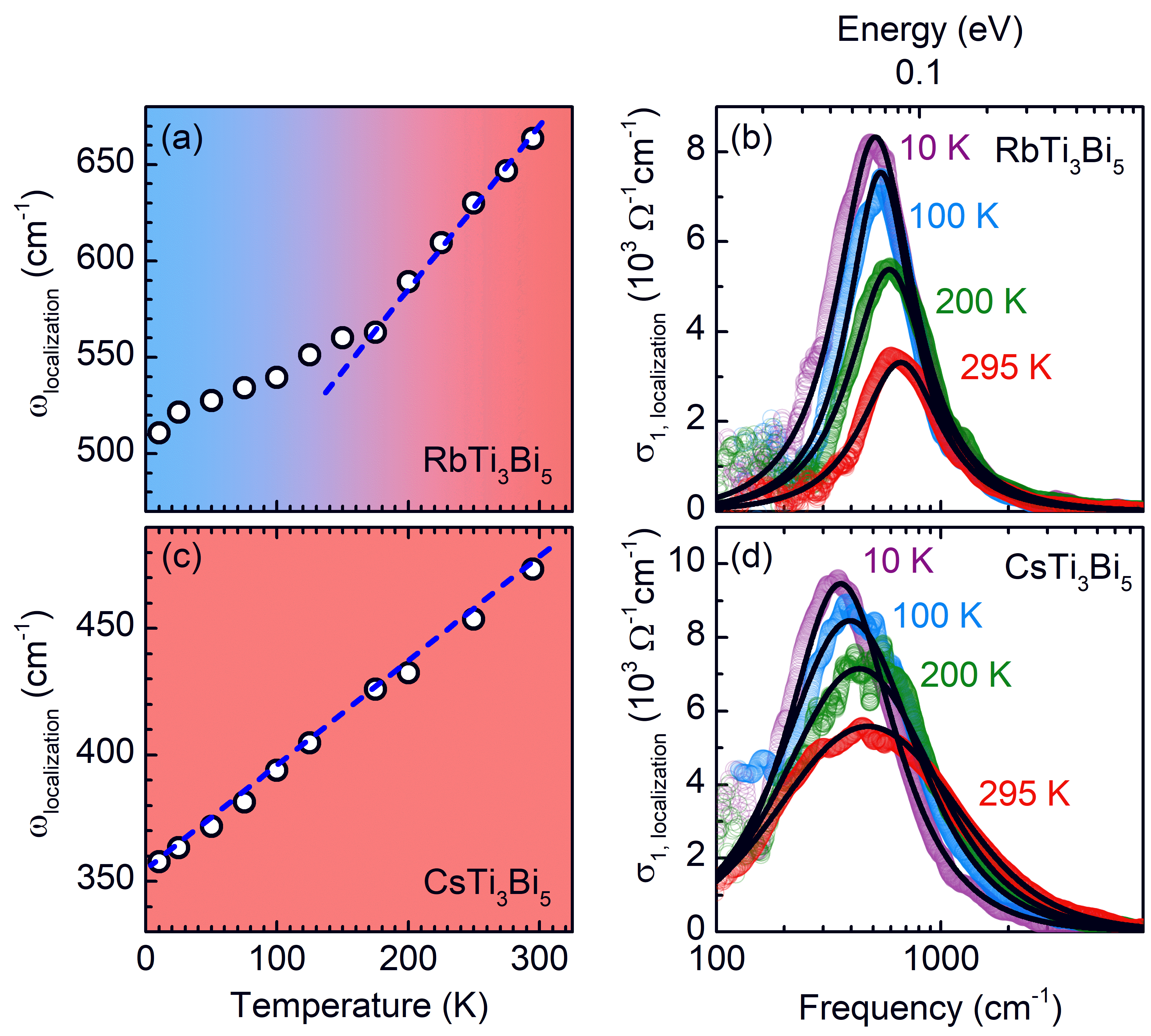}
\caption{(a) and (c) Temperature-dependent localization peak position in \RTB\ and \CTB, respectively. (b) and (d) Temperature evolution of the localization peak, obtained by subtracting the fitted Drude, interband contributions, and Fano resonance from the spectra. The solid lines are the fits of the localization peak from the decomposed optical conductivities given in Fig.~\ref{FigS3}.}
    \label{FigS4}
\end{figure}

In fact, a sharp absorption peak is in line with the theoretical framework by Fratini et $al.$, where a broader localization peak can evolve into a sharp resonance as the temperature reduces and the peak frequency approaches the frequency of the interacting bosonic mode \cite{Rammal2024}. Upon reducing the temperature, the localization peak in \RTB\ shows only a slight, continuous sharpening as illustrated in Fig.~\ref{FigS4}(b). 

The positions of the localization peaks as a function of temperature given in Figs.~\ref{FigS4}(a) and (c) for \RTB\ and \CTB, respectively, exhibit a linear red shift with decreasing temperature. In the case of \CTB, this trend continues down to the lowest temperatures, while for \RTB, a significant change in the slope is observed below $\sim$~150~K.

\section{Fano resonance}
\begin{figure}
\includegraphics[width=1\columnwidth]{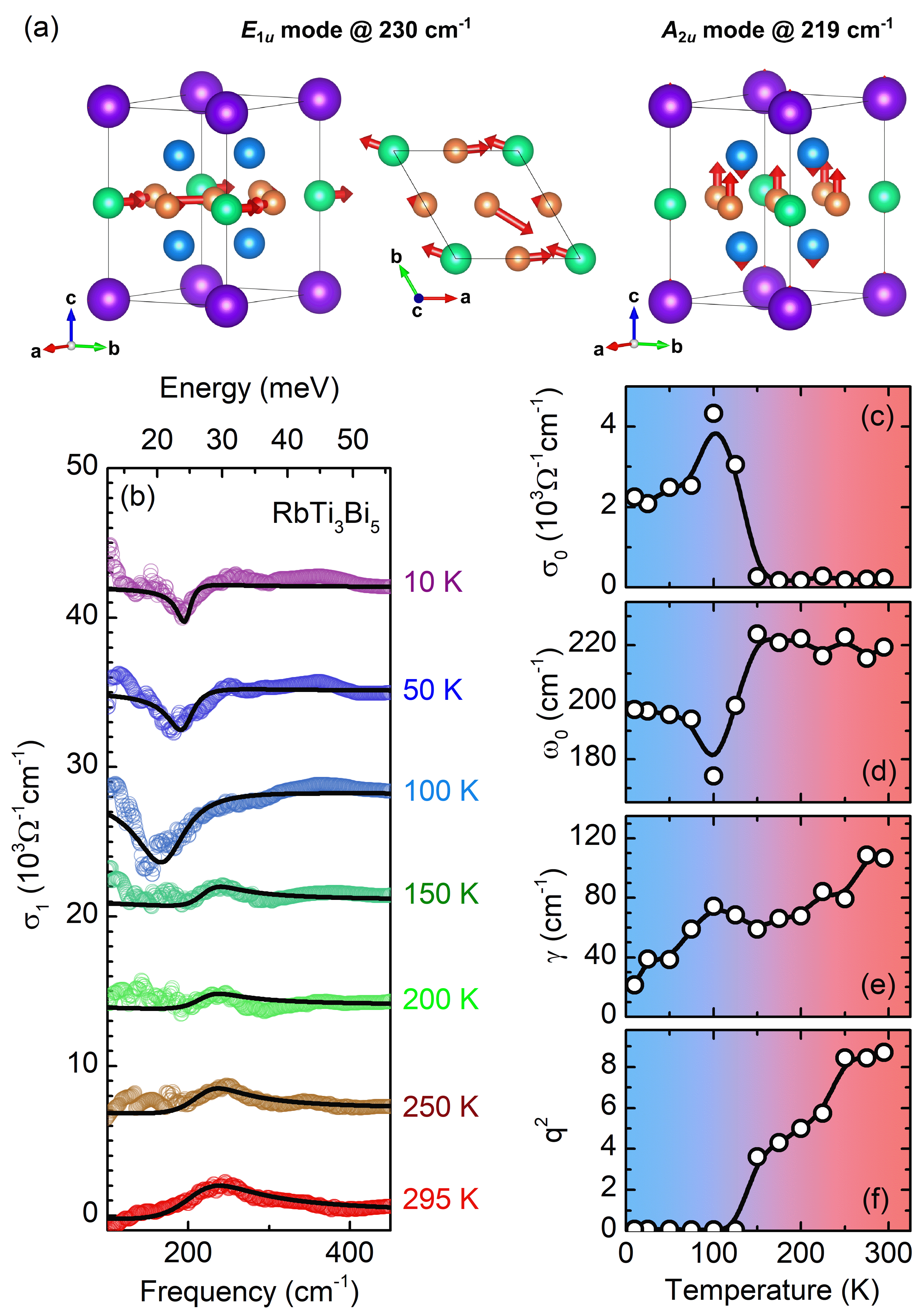}
\caption{(a) Sketch of the infrared-active in-plane $E_{\mathrm{1}u}$ (left) and out-of-plane $A_{\mathrm{2}u}$ (right) modes. (b) Optical conductivity at low energies presented as $\Delta \sigma_{\mathrm{1}}(\omega)$ with intra- and interband contributions subtracted and the baselines shifted by 7000~$\Omega^{-1}$\cm\ for clarity. The solid black lines are the fits to the phonon mode using the Fano model with the fit parameters given in panels (c-f).}
    \label{FigS5}
\end{figure}

$\Gamma$-point phonons were calculated in \texttt{VASP} \cite{Kresse1996} using the frozen-phonon method. These calculations reveal two infrared-active modes with energies comparable to the phonon mode observed in the room temperature spectrum of \RTB. One is an in-plane $E_{\mathrm{1}u}$ mode at 230~\cm\ involving mainly atomic displacements within the kagome Ti-Bi1 network as sketched in Fig.~\ref{FigS5}(a). The second mode is an out-of-plane $A_{\mathrm{2}u}$ mode at 219~\cm\ and, therefore, is not expected to be detected in our in-plane optical measurements.

Fig.~\ref{FigS5}(b) displays the temperature evolution of the phonon mode observed in \RTB, obtained after subtracting the intra- and interband contributions from the optical conductivity. Solid black lines are the Fano fits obtained by decomposing the optical spectra. Below 150~K a sudden change in the shape of the mode is observed. Here, the mode evolves into a sharp, more symmetric antiresonance, which is further accompanied by a redshift of the mode as seen in the fit parameters in panels (c-f).

\section{Additional DFT results}
Self-consistent density-functional calculations were performed on the $k$-mesh with 20~$\times$~20~$\times$~11 points. Optical conductivity was calculated on a denser $k$-mesh with up to 46~$\times$~46~$\times$~23 points. DFT+$U$ calculations were performed by adding a Hubbard $U$ to the Ti $d$-orbitals with the FLL (fully localized limit) and AMF (around mean field) double-counting corrections.

\begin{figure}
\includegraphics[width=1\columnwidth]{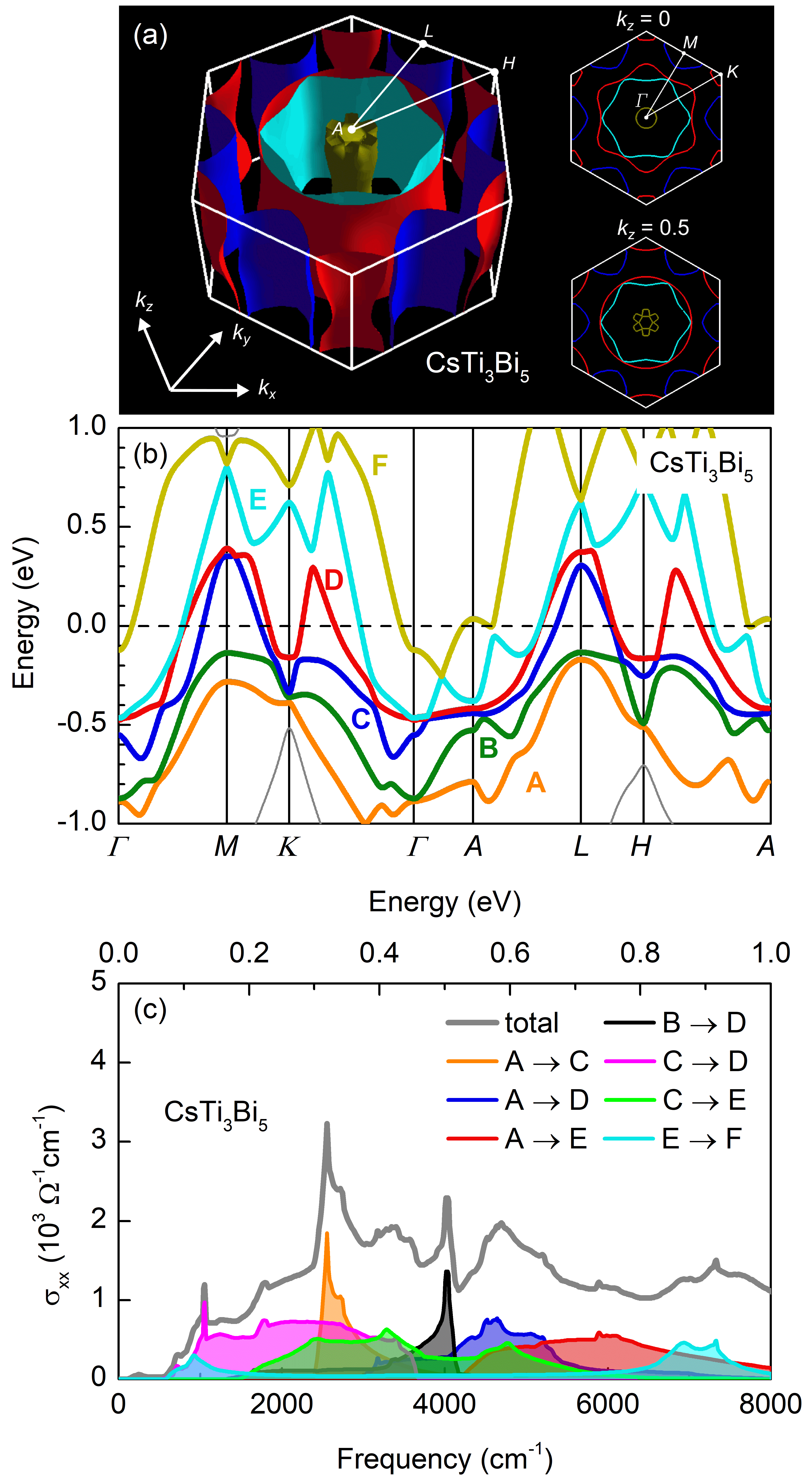}
\caption{(a) Fermi surface of \CTB\ constructed from the calculated band structure with the Fermi level shifted down by 179~meV to match the experimental data. \texttt{FermiSurfer} program was used for the visualization \cite{Kawamura2019}. (b) Calculated band structure of \CTB\ with the experimentally determined position of the Fermi level. (c) Band-resolved in-plane optical conductivity with different colors representing contributions from interband optical transitions between different bands following the labeling in panel (b). Note that no broadening is introduced to the calculations.}
    \label{FigS6}
\end{figure} 
\begin{figure}
\includegraphics[width=0.8\columnwidth]{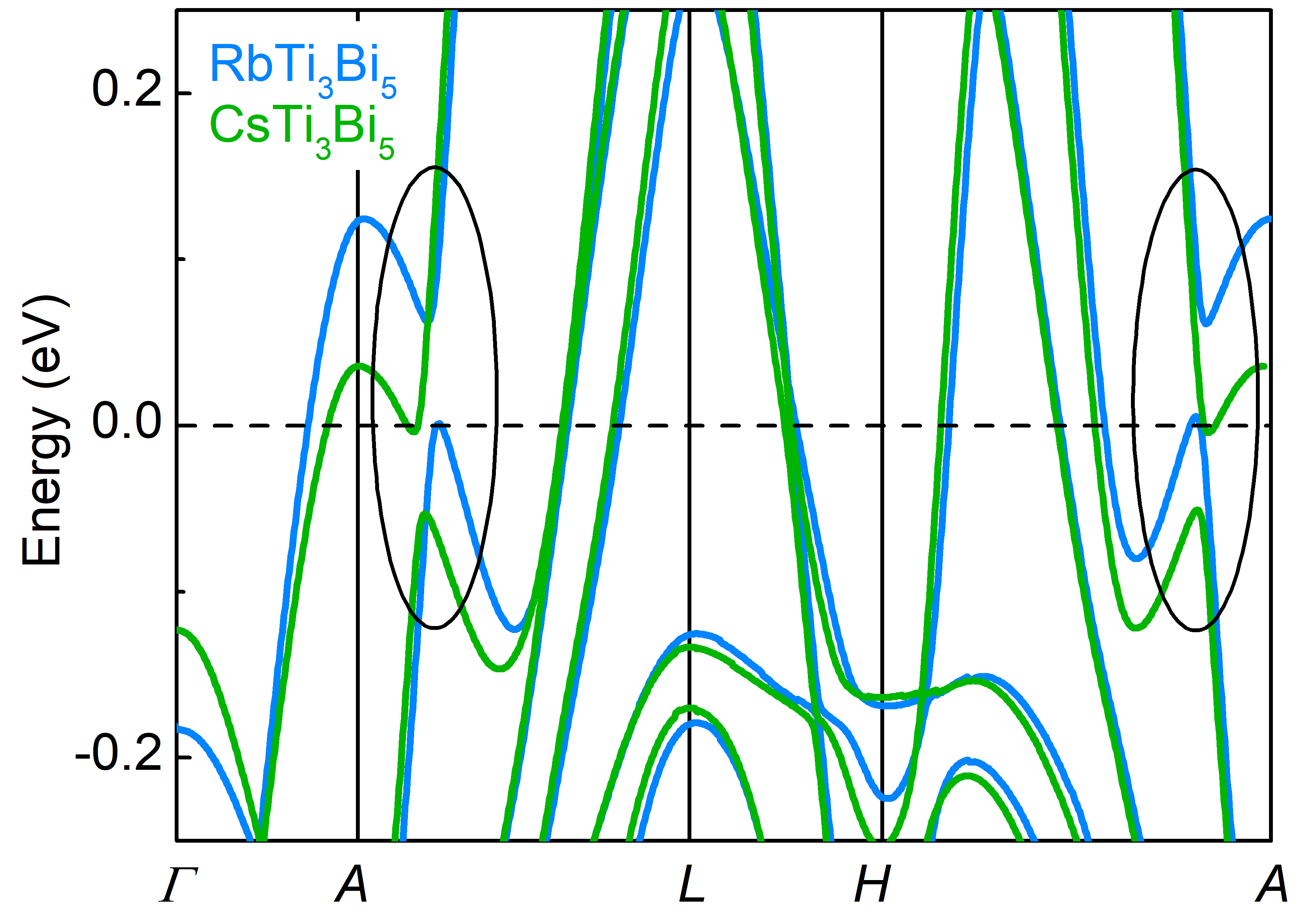}
\caption{Detailed view of the band structures at $k_z = 0.5$ for \RTB\ (blue) and \CTB\ (green). The Fermi levels are shifted downward as discussed in the main text. Black ovals mark the shift observed in the tilted Dirac-like bands along $A$-$L$ and $H$-$A$. }
    \label{bandsdetail}
\end{figure} 

\begin{figure}[h]
\includegraphics[width=1\columnwidth]{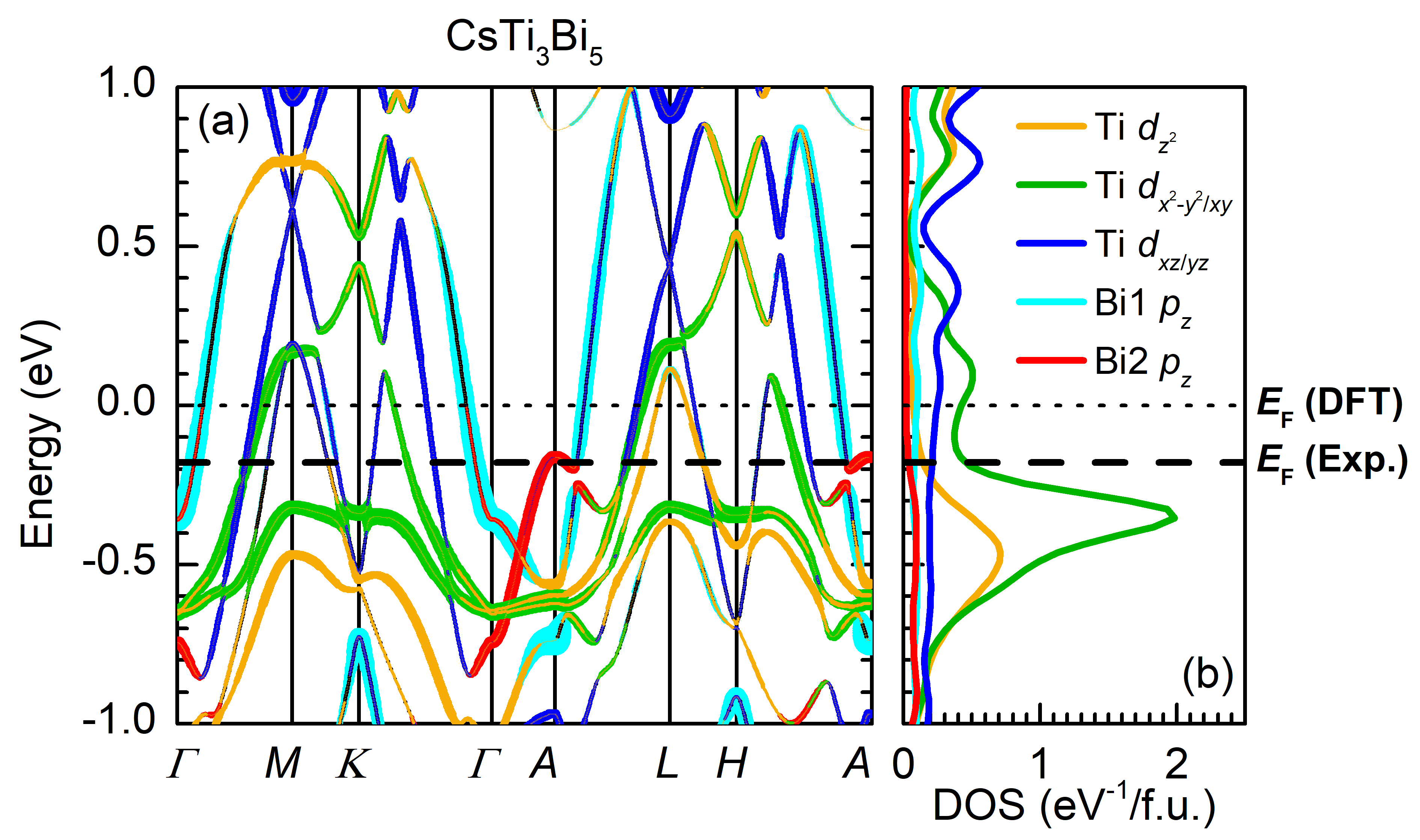}
\caption{(a) Band structure of \CTB\ with different colors representing contributions from different atomic orbitals as labeled in panel (b) showing the corresponding density of states (DOS). The experimentally determined position of the Fermi energy is shown by the dashed line. }
    \label{FigS7}
\end{figure}

Fig.~\ref{FigS6} presents the calculated Fermi surface, band structure, and the band-resolved optical conductivity of \CTB. In contrast to \RTB, the additional Fermi surface pockets at $k_z = 0.5$ are formed by band F according to the labeling in panel ~(b). Overall, the same interband contributions are observed as for \RTB\ [see Fig.~3(c) in main text], however, the low-energy transitions between band E and F are dampened, while transitions between band C and D have larger spectral weight at very low energies. As discussed in the main text, these fine differences are attributed to modifications of the tilted Dirac bands along $A$-$L$ and $H$-$A$. As presented in Fig.~\ref{bandsdetail}, the alkali ion leads to a shift of these Dirac points with respect to the Fermi energy, while the atomic contribution to the bands, namely Bi1 and Bi2 $p_z$ states, remains unaffected as illustrated in Fig.~\ref{FigS7} and Fig.~1(b) and (c).

\begin{figure*}
\includegraphics[width=2\columnwidth]{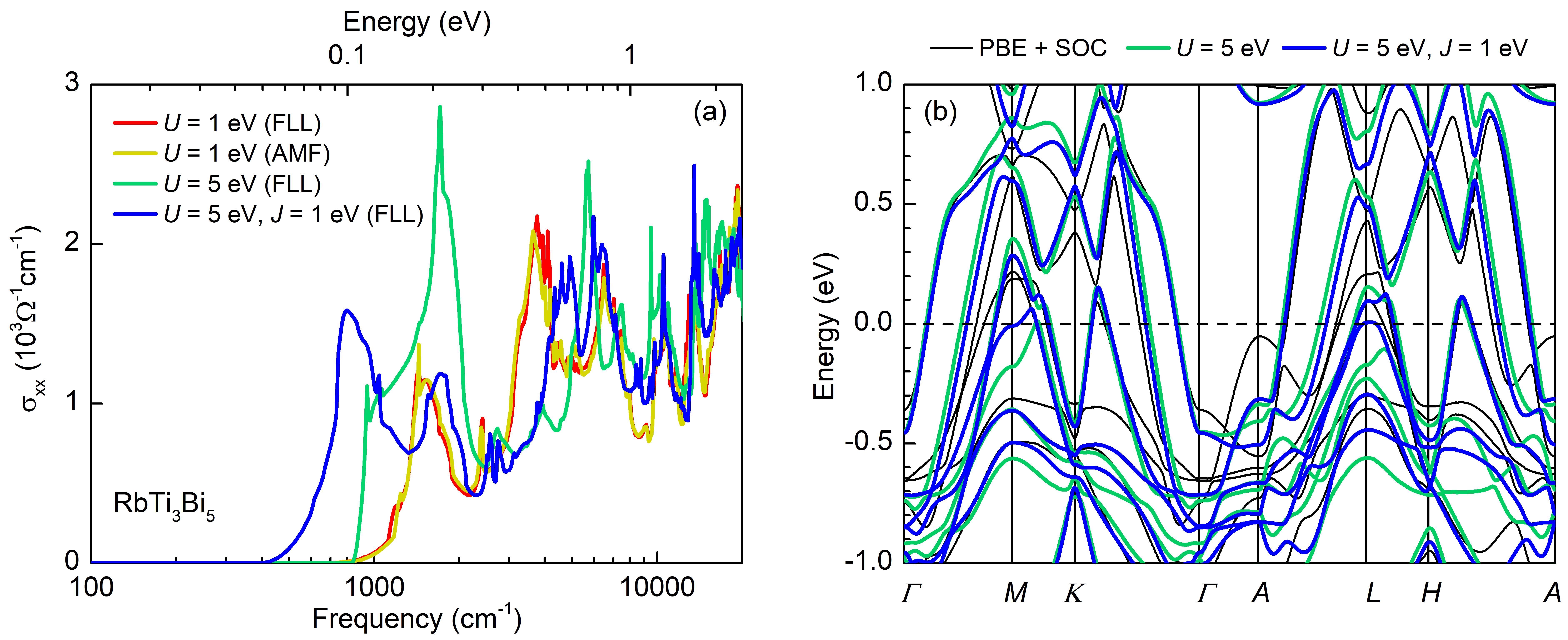}
\caption{(a) Calculated in-plane optical conductivities with different DFT+$U$ parameterizations. Note that no broadening is introduced to the calculations. While adding correlations to the Ti $d$-orbitals creates additional spectral weight at low energies similar to the experimental data, it significantly modifies the band structure as presented in panel (b). Most importantly, adding a Hubbard $U$ shifts the Ti flat bands away from the Fermi energy, contradicting the observations by ARPES studies. }
    \label{FigS8}
\end{figure*}

As discussed in the main text, the best agreement between the experimental and calculated optical conductivities is obtained upon shifting the Fermi level by -177~meV (\RTB) and -179~meV (\CTB). Nominally, this shift corresponds to a hole doping of -0.8$e$/f.u. (\RTB) and -0.9$e$/f.u. (\CTB) but it reflects the renormalization of band energies in the vicinity of the correlated flat bands, because such large changes in the number of electrons could not be caused by defects, impurities, and minor off-stoichiometry. Alternatively, we tried to reproduce the experimental results using DFT+$U$ calculations, but the agreement was poor, as shown in Fig.~\ref{FigS8}. While additional low-energy spectral weight is created by adding an on-site Coulomb repulsion $U = 5$~eV, this essentially shifts the Ti flat bands farther away from the Fermi energy, as well as leads to band saddle points below $E_{\mathrm{F}}$ [see panel (b) green line], which is inconsistent with the ARPES studies. We have further tested the inclusion of Hund's coupling with $J = 1$~eV shown in blue, resulting in a similar downward shift of the flat bands and significant changes in the Fermi surface at $k_z = 0$. 

\section{Plasma frequency}
The experimental plasma frequencies are obtained from the spectral weight (SW) analysis of the Drude and localization peaks according to
\begin{equation}
\omega_{\mathrm{p}} = \sqrt{\mathrm{SW_{Drude} + SW_{loc}}}.
\end{equation}
The spectral weight is calculated from the optical data by integrating the real part of the optical conductivity of the fitted Drude and localization peaks
\begin{equation}
\mathrm{SW} = \frac{1}{\pi^2\varepsilon_{\mathrm{0}}c} \int_0 ^{\omega_{\mathrm{c}}} \sigma_{\mathrm{1}}(\omega)\mathrm{d}\omega,
\end{equation}
with $c$ being the speed of light. The cut-off frequency is chosen as $\omega_{\mathrm{c}} = 50000$~\cm, taking into account the high-energy tail of the localization peak. 

Considering the error bars in determining the plasma frequency arising from (i) only catching the high-energy tail of the sharp Drude contribution in our measurements and (ii) the localization peak overlapping with interband transitions and the Fano resonance in the case of \RTB, the experimental plasma frequency does not depend on temperature. This further supports the absence of a structural phase transition that would likely create changes in the carrier density and/or the effective mass and consequently, in $\omega_{\mathrm{p}}$.

\FloatBarrier